\documentclass[twocolumn,amssymb,nobibnotes,aps,prbstab, reprint,letterpaper,pre/printnumbers,amsmath,
superscriptaddress,longbibliography,
showpacs,floatfix]{revtex4-1}
\usepackage{dcolumn}
\usepackage{bm}
\usepackage{graphics}
\usepackage{float}
\usepackage{epsfig}
\usepackage{multirow}
\usepackage{amsmath}
\usepackage{multirow}
\usepackage{tabularx}
\usepackage{bm}        
\usepackage{mathrsfs}
\usepackage{tabularx}
\usepackage{tabulary}
\usepackage{gensymb}
\usepackage{enumitem}
\usepackage[dvipsnames]{xcolor}
\usepackage[pass,paperwidth=8.5in,paperheight=11in]{geometry}

\usepackage[linktocpage=true]{hyperref}
\usepackage{hyperref}
\usepackage{comment}
\usepackage{soul}
\usepackage{notes2bib}
\hypersetup{
  pdfnewwindow=true, 
  colorlinks=true,
  linkcolor=blue, 
  anchorcolor=blue,
  citecolor=blue, 
  filecolor=blue,
  menucolor=blue, 
  urlcolor=blue}

\usepackage{gensymb}
\usepackage{array}
\usepackage[normalem]{ulem}
\usepackage{xcolor}
\usepackage{orcidlink}

\begin{document}


\title{Engineering Topology by Design in Two-dimensional Materials}

\author{Arjyama Bordoloi\,\orcidlink{0009-0006-2760-3866}}
\affiliation{Department of Mechanical Engineering, University of Rochester, Rochester, New York 14627, USA}

\author{Sobhit Singh\,\orcidlink{0000-0002-5292-4235}}
\email{s.singh@rochester.edu}
\affiliation{Department of Mechanical Engineering, University of Rochester, Rochester, New York 14627, USA}
\affiliation{Materials Science Program, University of Rochester, Rochester, New York 14627, USA}

\begin{abstract}
Two-dimensional topological insulators (2D TIs) have emerged as a cornerstone of next-generation spintronic technologies due to their robust, dissipationless edge states protected by time-reversal symmetry. Initial realizations of 2D TIs have primarily focused on materials with strong intrinsic spin–orbit coupling capable of driving band inversion, an approach that significantly constrains the accessible materials landscape. More recently, a paradigm shift has occurred toward engineering topological phases in van der Waals (vdW) heterostructures, where nontrivial band topology can arise from interfacial coupling rather than relying solely on intrinsic material properties. This framework provides an exceptionally versatile platform with multiple tunable degrees of freedom, including stacking configuration, twist angle, and chemical functionalization, allowing systematic manipulation of the band topology. Furthermore, external stimuli, such as electric fields, strain, and light-matter coupling, enable dynamic and reversible control of the topological character. The combined use of vdW interface engineering and external modulation allows the realization of 2D TI phases even in otherwise topologically trivial systems, substantially expanding the accessible materials landscape. This Research Update reviews key milestones in the development of vdW-engineered 2D topological quantum materials, critically assesses outstanding theoretical and experimental challenges, and outlines promising directions for future breakthroughs.
\end{abstract}

\maketitle
\section{Introduction}

\begin{figure*}
\centering
\includegraphics [width=17cm]{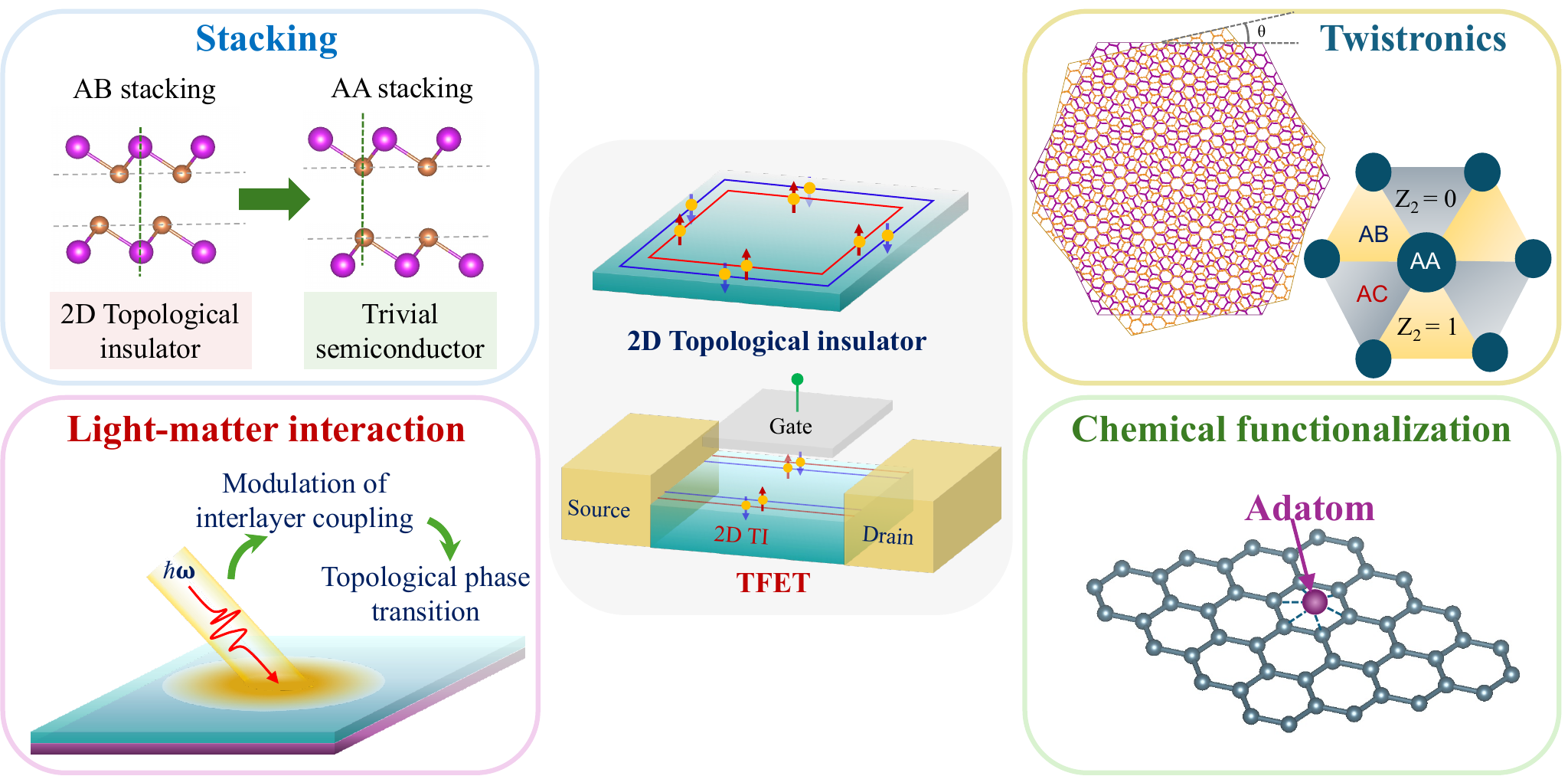}
\caption{Schematic illustration of multiple pathways to induce topological phase transitions in vdW materials. Top left: different stacking arrangements lead to distinct topological phases. Top right: local stacking configurations (AA, AB, AC) within a moiré supercell correspond to regions with different topological character(Z$_2$\,=\,1, topologically nontrivial; Z$_2$\,=\,0, trivial).
Center: topological field-effect transistor (TFET) implementation. Bottom left: terahertz light–matter interactions stabilize metastable topological states by modifying electronic properties. Bottom right: chemical functionalization tunes topology via SOC and charge transfer.}
\label{fig:phase diagram}
\end{figure*}

The discovery of the quantum Hall (QH) effect in 1980~\cite{klitzing_PRL_1980} marked the beginning of a topological revolution in condensed matter physics. Unlike conventional Landau-Ginzburg theory~\cite{ginzburg_Physik_journal_2009}, which classifies phases based on symmetry breaking, the QH effect introduced a topological classification of materials~\cite{konig_Journal_of_the_phy_soc_japan_2008quantum, Hasan_RevModPhys_2010, Qi_RevModPhys_2011}. It arises in systems subjected to extremely strong magnetic fields, resulting in quantized edge conductance while the bulk remains insulating. Electrons in these chiral edge states propagate unidirectionally, suppressing backscattering and hence enabling dissipationless transport. In the QH effect, the Hall conductance is precisely quantized in integer multiples of $e^2/h$ which is associated with a topological invariant known as the Chern number~\cite{Laughlin_PRB1981, thouless_PRL_1982}.

However, the strong magnetic field required in conventional QH systems breaks time-reversal symmetry (TRS). In the quest to realize topological phases that preserve TRS, Haldane introduced a key breakthrough: he demonstrated that a QH-like state can emerge in the absence of any net magnetic field, which was realized via complex next-nearest-neighbor hopping on a honeycomb lattice~\cite{Haldane_PRL_1988}.
Subsequently, strong spin–orbit coupling (SOC) was shown to mimic the role of an effective magnetic field while preserving TRS, leading to the prediction of the quantum spin Hall (QSH) phase, also known as two-dimensional topological insulators (2D TIs)~\cite{kane_PRL_2005, bernevig_Science_2006}. 2D TIs were first theoretically predicted in 2006~\cite{bernevig_Science_2006} and subsequently realized experimentally in HgTe/CdTe quantum wells, where a topological phase transition occurs upon varying the well thickness~\cite{konig_Science_2007, roth_Science_2009}. In these systems, electrons propagate along helical edge states, where electrons of opposite spin move in opposite directions, forming a “two-lane highway” for spin-polarized transport. 

Owing to their dissipationless edge transport, 2D TIs offer a wide range of topology-enabled quantum device applications, providing a potential alternative to conventional semiconductor spintronic devices with higher energy efficiency and reduced heat dissipation~\cite{Murakami_Science_2003, zhang_Small_2022, jin_Nanoscale_2023, gilbert_Communication_Phy_2021, kou_Journal_of_phy_chem_c_2017}. They not only introduce a new means to control the electronic properties of conventional electronic, optoelectronic, spintronic, and memory devices, but also offer enhanced precision and robustness~\cite{tian_Materials_2017, gilbert_Communication_Phy_2021, jin_Nanoscale_2023, zhang_Small_2022}. Since the helical edge states are protected by TRS, they are immune to nonmagnetic impurities and structural defects, resulting in pure, quantized spin currents~\cite{Chang_Science_2013, konig_Journal_of_the_phy_soc_japan_2008quantum, Hasan_RevModPhys_2010, Qi_RevModPhys_2011}. The spin-momentum locked edge states are free from electron backscattering, further minimizing energy loss. Moreover, when TIs are placed in proximity to superconducting materials, they can host Majorana bound states, opening a pathway toward fault-tolerant topological quantum computing~\cite{fu_PRL_2008, kitaev_Physics_uspekhi_2001, lutchyn_PRL_2010, sarma_npj_quantum_information_2015, beenakker_Annual_rev_condensed_mat_phy_2013, leijnse_Semiconductor_science_and_technology_2012, fu_PRL_2008}.


Driven by both the growing demand for application-oriented aspects and the pursuit of a deeper understanding of fundamental physics, significant progress has been made over the past two decades in designing and discovering novel topological quantum materials with tailored properties~\cite{hasan_Review_of_modern_phy_2010, kou_Journal_of_phy_chem_c_2017, qi_rev_modern_phy_2011, Singh_npjQM2022, anirban_Nature_review_phy_2023, ni_Chemical_reviews_2023}. A key requirement for realizing time-reversal-invariant topological phases in 2D materials is strong SOC, which can induce a band inversion near the Fermi level~\cite{konig_Science_2007, hsieh_Science_2009, fu_PRB_2007, fu_PRL_2007, moore_PRB_2007, Bordoloi_PRB_2024}. 
Nevertheless, even materials with substantial SOC can remain topologically trivial~\cite{Singh_PRB_2017, ishizaka_Nature_materials_2011, bordoloi_2D_Materials_2025}. 
Moreover, achieving materials with robust and tunable topological edge states remains challenging, restricting the available materials landscape. Consequently, there has been renewed interest in rationally engineering topological materials using various design strategies~\cite{Das2013_nature, bordoloi_2D_Materials_2025, Nechaev2017_sci_rep, kou_APL_2018, bao_Nature_Rev_Phy_2022}. 
These approaches aim either to convert trivial materials into topologically nontrivial phases or to enable tunable topological phase transitions, providing control over the transition between distinct topological states~\cite{sanchez2025switching, Das2013_nature, Nechaev2017_sci_rep, g_MoS2_bookchapter2019, bordoloi_2D_Materials_2025, tong_Nature-phy_2017, Singh_PRB2016}.

To this end, two-dimensional vdW materials provide a highly versatile platform due to their reduced dimensionality, rich structural, electronic, mechanical, and chemical degrees of freedom, and the tunability of their properties via external stimuli such as electric fields, strain, optical excitation, and chemical functionalization~\cite{zhang_Small_2022, duong_ACS_nano_2017, Singh_JPCL2019, novoselov_Science_2016, liu_Nature_review_materials_2016, SharmaPRB2025}. 2D vdW materials consist of atomically thin layers with strong intralayer covalent bonding and weak interlayer van der Waals interactions. The weak interlayer coupling facilitates exfoliation down to the monolayer limit and allows for significant structural flexibility. Further, the anisotropic bonding environment enables the realization of different stacking arrangements through relative sliding of one layer with respect to another~\cite{li_Advanced_Materials_2024, han_Nano_Letters_2025}. Additionally, controlling the relative orientation (twist angle) between layers enables moiré engineering, giving rise to a wide range of exotic electronic phenomena, including 2D Mott insulators~\cite{mak_Nature_Nanotechnology_2022, li_Nature_Nanotechnology_2021, regan_Nature_2020} and Wigner crystals~\cite{regan_Nature_2020, matty_Nature_com_2022, padhi_PRB_2021, li_Nature_2021, nuckolls_Nature_Review_materials_2024}. Beyond structural manipulation, chemical functionalization, such as intercalation~\cite{yang_Nature_Review_Chemistry_2024, li_Small_methods_2021}, adatom adsorption~\cite{Brey_PhysRevB_2015, Weeks_PhysRevX._2011, Zhang_PhysRevB_2013}, or surface modification~\cite{reed_Advanced_materials_interfaces_2022, mozafari_Materials_advances_2021}, provides further control, allowing the tailoring of electronic, optical, and magnetic properties in these materials.

In this Research Update, we summarize some of the key design principles for engineering tunable topological phases in 2D vdW materials. We focus primarily on four strategies: 
\begin{itemize}[itemsep=0.5pt] 
    \item Stacking
    \item Twisting
    \item Light–matter interactions
    \item Chemical functionalization
\end{itemize}

Stacking and twisting modulate the relative atomic arrangements and interlayer coupling between constituent layers, which significantly renormalizes the electronic band structure and, consequently, the band topology. Stacking is particularly noteworthy, as it has been explicitly demonstrated to transform topologically trivial constituents into an overall topologically nontrivial system~\cite{Das2013_nature, bordoloi_2D_Materials_2025, Nechaev2017_sci_rep}. Twisting, on the other hand, can produce spatially modulated topological domains\,--\,a so-called topological mosaic pattern\,--\,within the moiré supercell~\cite{tong_Nature-phy_2017}.

Light–matter interactions, though traditionally employed for spectroscopic characterization, have recently emerged as a powerful tool for the dynamic manipulation of topological phases~\cite{qi_ACS_nano_2022, vaswani_PRX_2020, zhang_PRX_2019, luo_Nature_Materials_2021, bao_Nature_Rev_Phy_2022}. By coupling to either electronic or lattice degrees of freedom, light can drive systems into nonequilibrium phases with metastable states that exhibit distinct structural and topological properties. On the other hand, chemical functionalization remains one of the most extensively explored strategies for engineering topological phases, starting from prototypical 2D materials such as graphene~\cite{Brey_PhysRevB_2015, Weeks_PhysRevX._2011}. Finally, we discuss emerging device applications of engineered 2D TIs, including topological field-effect transistors (TFETs), memory devices, and topological p–n junctions. In the Summary and Outlook section, we highlight outstanding challenges in the field and provide perspectives on promising future directions. 
Overall, this Research Update emphasizes the rational engineering of tunable topological phases in 2D vdW systems, in contrast to conventional approaches that focus primarily on discovering materials with intrinsic topological character.

\section{Engineering strategies}
This section summarizes the various design strategies for engineering tailored topological phases, including stacking, twisting, light–matter interactions, and chemical functionalization. Each subsection focuses on a specific design principle, providing a conceptual overview at the outset, followed by a summary of recent advancements in the field.

\subsection{Stacking}

A highly promising route to engineer exotic quantum phases in layered vdW materials is the deliberate modulation of their stacking configuration, which refers to the relative lateral displacement or sliding between adjacent constituent layers. vdW materials are characterized by strong intralayer covalent bonding coexisting with weak interlayer interactions~\cite{geim_Nature_2013, castellanos_Nature_Reviews_2022, liu_Nature_review_materials_2016, novoselov_Science_2016}. Owing to this pronounced anisotropy in bonding strength, vdW systems provide an exceptionally fertile platform for realizing reversible, low-energy, multistate transitions between distinct stacking configurations, with energy barriers that are orders of magnitude smaller than those required to break conventional covalent or ionic bonds~\cite{fox_Chemical_reviews_2023, yang_Journal_physical_chem_letters_2018, wei_arxiv_2025, li_ACS_Nano_2017, Xiao_PRL_2018, Singh_PRL_2020, wu_PNAS_2021, Huang_NatComm2019}.

Variations in stacking order modify the atomic registry between layers, which in turn governs the coupling between multiple internal degrees of freedoms, including crystal symmetry, polarization, orbital hybridization, and interlayer coupling strength. This strong interplay between stacking configuration and electronic structure enables a programmable switching mechanism that can toggle between distinct quantum phases within the same material~\cite{bordoloi_2D_Materials_2025, kou_APL_2018, wei_arxiv_2025, kou_APL_2018}. Since the energy barriers associated with interlayer sliding are relatively low, stacking transitions can be readily induced using experimentally accessible static or dynamic stimuli, such as mechanical strain~\cite{oviedo_ACS_Nano_2015, fox_Chemical_reviews_2023}, optical excitation~\cite{wei_arxiv_2025, vaswani_PRX_2020, zhang_PRX_2019, luo_Nature_Materials_2021, bao_Nature_Rev_Phy_2022}, electrostatic gating~\cite{bordoloi_2D_Materials_2025}, pressure~\cite{Liang_SciAdv2017, li_Nature_materials_2019, song_Nature_Materials_2019, Karki2022}, or chemical functionalization~\cite{rajapakse_npj_2d_mat_applications_2021, muscher_Advanced_materials_2021, wang_Advanced_materials_2022}.

In recent years, alongside the exploration of stacking-controlled phenomena such as sliding ferroelectricity~\cite{wu_PNAS_2021, vizner_Science_2021l, han_Nano_Letters_2025, yang_Advanced_functional_materials_2023}, interlayer exciton polarization~\cite{ciarrocchi_Nature_photonics_2019, jiang_NaturE_nanotech_2021, wilson_Nature_2021}, piezophotovoltaic effects~\cite{aftab_Advanced_Optical_Materials_2022}, magnetism~\cite{Liu_PRL_2020, song_Nature_Materials_2019, Liu_NL2019, Liu_NL2021, Ren_PRB_2022, huang_Nature_2017}, superconductivity~\cite{knobel_Advanced_materials_interface_2025}, and multiferroicity~\cite{xun_Nano_Letters_2024, pan_APL_2025, chen_Nature_comm_2025}, there has been a rapid surge of interest in exploiting stackingtronics as a powerful route to realize and manipulate emergent topological properties in vdW materials. 

Stacking-induced topological phase transitions can arise through multiple mechanisms. First, variations in stacking configuration modify the relative atomic registry between adjacent layers, leading to substantial changes in steric intercations via orbital overlap and interlayer hybridization~\cite{bordoloi_2D_Materials_2025, kou_APL_2018, li_npj_computational_material_2025}. Such modifications can significantly reconstruct the electronic band structure, thereby driving topological phase transitions. Second, the relative lateral displacement between layers governs the global crystal symmetry of the system, even when the symmetry of individual monolayers remains unchanged~\cite{Singh_PRL_2020}. Because topological phases are closely linked to specific symmetries, stacking transitions effectively act as discrete symmetry switches, capable of breaking or restoring key symmetries and directly controlling the topological character. Furthermore, in vdW heterostructures, stacking-dependent proximity effects, including modulation of spin-orbit coupling strength~\cite{kou_ACS)nano_2014, gmitra_PRB_2015, jin_PRB_2013, g_MoS2_bookchapter2019, Singh_PRB_2018}, magnetic exchange interactions~\cite{lee_Nature_comm_2016, li_PRB_2015, bhattacharyya_Advanced_materials_2021}, and proximity-induced superconductivity~\cite{black_PRB_2013, khanna_PRB_2014, sau_PRB_2010, huang_ACS_Nano_2018, kezilebieke_Nature_2020}, can further modify the topological properties of constituent layers.

Below, we summarize recent progress in this area, categorized into two parts: (1) Induced topological phase transitions in homobilayers, where the constituent layers are of the same material, and (2) Transitions in heterostructures driven by proximity effects, where the constituent layers are composed of different materials.

\subsubsection{Stacking-induced nontrivial topological phases in homobilayers}

\begin{figure}[!!t]
\centering
\includegraphics [width=1\columnwidth]{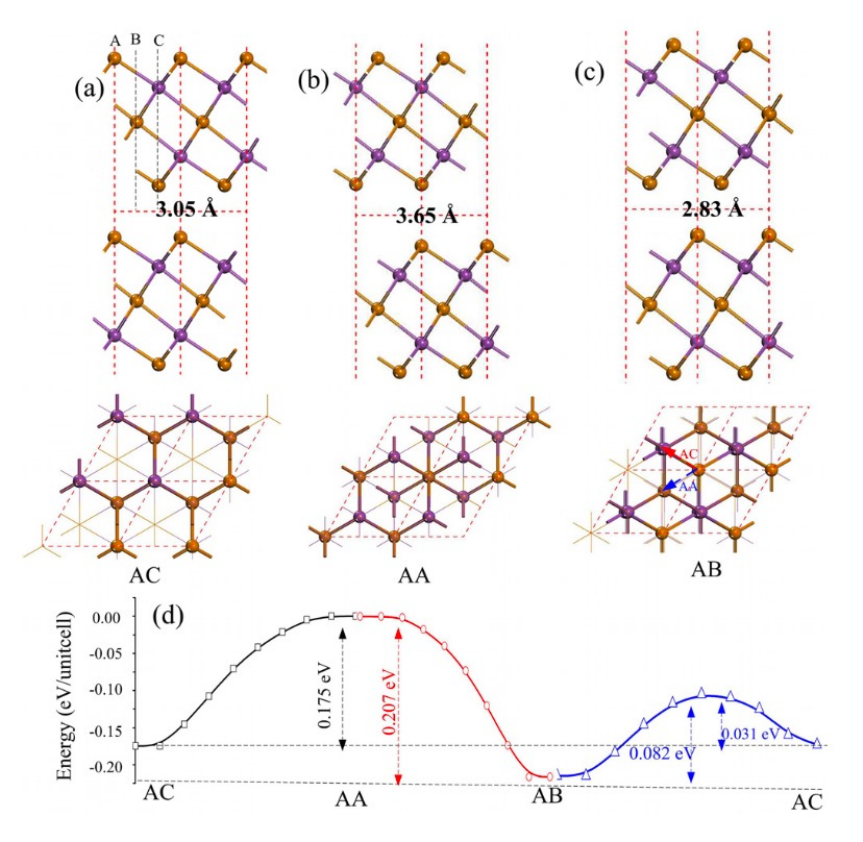}
\caption{Side and top views of Bi$_2$Te$_3$ bilayers in (a) AC, (b) AA, and (c) AB stacking configurations. Purple and yellow spheres represent Bi and Te atoms, respectively. (d) Energy barriers associated with switching between different stacking configurations, calculated using nudged elastic band (NEB) methods (Figure adapted with permission from Ref.~\cite{kou_APL_2018}).}
\label{fig:stacking_Bi2Te3}
\end{figure}

Bi$_2$Te$_3$ is a prototypical two-dimensional topological insulator that supports multiple stacking configurations with distinct interlayer bonding features, leading to stacking-controlled topological phase switching~\cite{kou_APL_2018}. Among the high-symmetry stacking arrangements\,-\,AA, AB, and AC\,-\, (Figure~\ref{fig:stacking_Bi2Te3}) the AB stacking is energetically most stable and exhibits a nontrivial topological phase, whereas the AA and AC stackings are topologically trivial. Notably, reversible transitions between these stacking configurations occur across a relatively low energy barrier of 30–80 meV per unit cell, as illustrated in Fig.~\ref{fig:stacking_Bi2Te3}.
In addition, an oscillatory crossover between topologically trivial and nontrivial phases as a function of thickness has been reported in Bi$_2$Te$_3$/Bi$_2$Se$_3$ heterostructures~\cite{Liu_PRB_2010, lu_PRB_2017}, and subsequently verified experimentally by Kim \textit{et al.}~\cite{Kim_PRB_2011}.

MnBi$_2$Te$_4$ has been theoretically predicted to undergo a stacking-induced topological phase transition between Chern insulating phases with $C$=1 and $C$=0, driven by a lateral shift of the topmost septuple layer~\cite{li_npj_computational_material_2025, Varnava2021}.
Similarly, for SnTe stacked along the [001] direction~\cite{araujo_Scientific_Reports_2018}, an even or odd number of monolayers results in a symmorphic or nonsymmorphic space group, respectively. In symmorphic stackings, hybridization between surface states can induce band inversion and topological phase transitions, whereas nonsymmorphic stacking enforces additional symmetry-protected degeneracies that suppress surface-state hybridization and prevent topological transitions~\cite{araujo_Scientific_Reports_2018}.

\begin{table}[!!b]
\centering
\renewcommand{\arraystretch}{1.2} 
\caption{Reported bandgaps for various vdW homobilayers exhibiting stacking-induced nontrivial topological phase transitions}
\label{tab:stacking_TI}
\begin{tabular}{l c c}
\hline
\textbf{Material system} & \textbf{Bandgap} & \textbf{Reference} \\
\hline
Bi$_2$Te$_3$ & 0.03 eV & \cite{kou_APL_2018} \\
GeBi$_2$Te$_4$  & 0.09 eV & \cite{peng_PRB_2020} \\
SnBi$_2$Te$_4$  & 0.06 eV & \cite{peng_PRB_2020} \\
PbBi$_2$Te$_4$  & 0.10 eV & \cite{peng_PRB_2020} \\
BiSb-SbBi  & Semimetallic & \cite{bordoloi_2D_Materials_2025} \\
Ge$_2$Bi$_2$Te$_5$  & 0.007 eV & \cite{Li_PhysRevB_2023_stacking} \\
Sn$_2$Bi$_2$Te$_5$ & 0.046 eV & \cite{Li_PhysRevB_2023_stacking} \\
Pb$_2$Bi$_2$Te$_5$  & 0.112 eV & \cite{Li_PhysRevB_2023_stacking} \\
OsClBr  & Small bandgap semiconductor & \cite{li_APL_2023strain} \\
\hline
\end{tabular}
\end{table}

Beyond material-specific realizations, recently stacking has also been proposed as a general strategy to engineer TIs from systems that are intrinsically topologically trivial. In 2013, Das \textit{et al.}~\cite{Das2013_nature}, using an effective Hamiltonian model, introduced a novel approach of designing a 3D TI by stacking up bilayers made of two-dimensional Fermi gases with opposite Rashba spin-orbit coupling (RSOC) on adjacent layers. By adjusting various parameters such as the Rashba parameter, intra-layer coupling (Dirac mass, $D_0$), and inter-layer coupling (t$_z$) within practically realizable limits, they observe that while a single bilayer exhibits topologically trivial behavior, topologically nontrivial insulating states emerged after reaching a critical number of bilayers. Extending this model, Gunnink \textit{et al.}~\cite{Gunnink_2020_JPCM} showed that a single bilayer can become topologically nontrivial when the intralayer coupling ($D_0$) is negative. Furthermore, successive stacking of such bilayers gives rise to distinct topological phases, determined by $D_0/t_z$ ratio. In a related material realization, a centrosymmetric sextuple layer formed by two BiTeI trilayers with opposite RSOC was shown to host a sizable inverted band gap, making it a promising platform for practical topological applications~\cite{Nechaev2017_sci_rep}.

\begin{figure*}[hbtp]
\centering
\includegraphics [width=16cm]{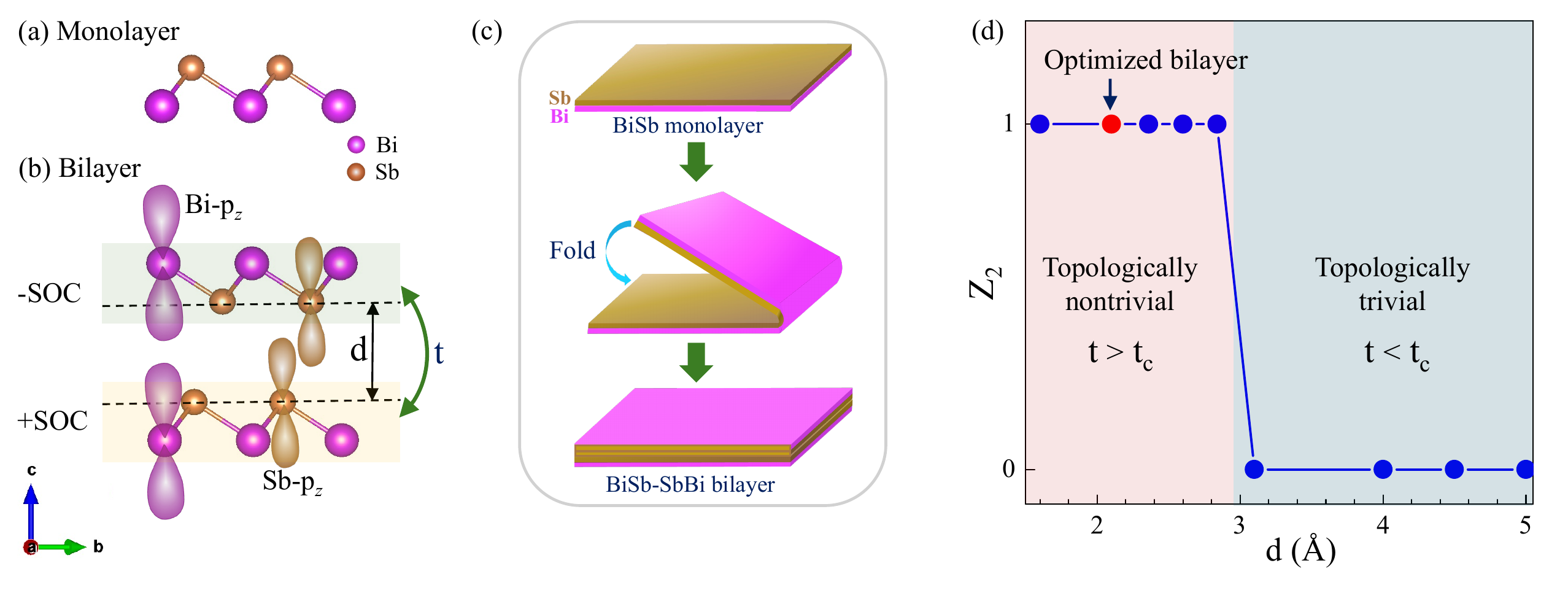}
\caption{Optimized crystal structures of (a) BiSb monolayer and (b) BiSb–SbBi bilayer. Here, d and t denote the interlayer separation and electron tunneling, respectively. (c) Schematic of a possible experimental route to realize the BiSb–SbBi bilayer. (d)\,Topological phase transition as a function of d, with the light red region indicating Z$_2 = 1$ and the light blue region indicating Z$_2 = 0$ (Figure adapted from Ref.~\cite{bordoloi_2D_Materials_2025}).}
\label{fig:stacking_BiSb}
\end{figure*}

Similarly, the BiSb–SbBi bilayer~\cite{bordoloi_2D_Materials_2025}, formed by stacking two BiSb monolayers in an inverted configuration as shown in Fig.~\ref{fig:stacking_BiSb}, exhibits a topologically nontrivial phase with Z$_2$=1. While the monolayer exhibits topologically trivial behaviour despite having strong spin–orbit coupling~\cite{Singh_PRB_2017}, interlayer hybridization between Bi-p$_z$ orbitals induces nontrivial behavior in the bilayer. The topological character is fundamentally governed by a delicate interpaly between spin–orbit coupling strength and interlayer electron tunneling. An applied electric field further enables dynamic control over the system's topology, paving the way for gate-tunable topological phases~\cite{bordoloi_2D_Materials_2025}.

In the same spirit, MBi$_2$Te$_4$\,(M = Ge, Sn, Pb)~\cite{peng_PRB_2020} is a topologically trivial semiconductor in its monolayer form, whereas the bilayer exhibits stacking-dependent topological phases due to strong hybridization of Te-$p_z$ orbitals at the interface. Further, single-layer M$_2$Bi$_2$Te$_5$\,(M\,=\,Ge, Sn, Pb) is a topologically trivial indirect band-gap semiconductor, but bilayer M$_2$Bi$_2$Te$_5$\,(M\,=\,Ge, Sn, Pb) can become topologically nontrivial with a sizable band gap of 111.6 meV depending on the stacking order~\cite{Li_PhysRevB_2023_stacking}. OsClBr, on the other hand, is a dipolar ferromagnetic semiconductor that is trivial in the monolayer form, yet exhibits a valley quantum anomalous Hall phase in a specific bilayer stacking configuration, which remains robust under applied strain~\cite{li_APL_2023strain}.

\subsubsection{Proximity-induced topological phase transitions in heterostructures}

Proximity effects in vdW heterostructures provide an effective route to realizing nontrivial topological phase transitions. A notable example is graphene, whose intrinsically weak SOC can be substantially enhanced through proximity to materials with strong SOC, thereby stabilizing topological phases against thermal fluctuations and enabling room-temperature applications~\cite{gmitra_PRB_2015, jin_PRB_2013, wang_Nat_comm_2015, Singh_PRB_2018}. When graphene is epitaxially grown on the topological insulator Sb$_2$Te$_3$, proximity-induced SOC opens a sizable gap of approximately 20 meV~\cite{jin_PRB_2013}. Similarly, at the interface between monolayer graphene and few-layer WS$_2$, the induced SOC strength reaches about 17 meV~\cite{wang_Nat_comm_2015, avsar_Nat_comm_2014}. Furthermore, graphene sandwiched between BiTeX (X = Br, Cl, I) layers, although BiTeX itself is not a topological insulator, can host a two-dimensional topological insulating phase due to the strong SOC of the BiTeX layers~\cite{kou_ACS)nano_2014}. In this case, a band gap of up to 70 meV can be achieved, which can be further enhanced under applied pressure.

Apart from graphene-based systems, several other vdW heterostructures have also been shown to host proximity-induced topological phases. For instance, silicene grown on 4H-SiC has been reported to exhibit a topologically nontrivial phase, with the realization of a quantum anomalous Hall state in the resulting heterostructure~\cite{Krawiec_JPCM_2018}. Monolayer Pb(111) has also been predicted to form a two-dimensional Z$_2$ topological insulator when interfaced with NbSe$_2$, driven by strong hybridization between the Pb and NbSe$_2$ electronic states~\cite{Huang_PhysRevB_2014}. Furthermore, $\beta$-antimonene, which is a trivial insulator in its pristine form, can acquire a symmetry-protected spin texture from the underlying substrate via proximity effects, leading to an outward migration of topological states~\cite{holtgrewe_scientific_reports_2020}. Experimental angle-resolved photoemission spectroscopy (ARPES) measurements have further established that the Sb$_2$Se$_3$/Bi$_2$Se$_3$ heterostructure behaves as a topological insulator, whereas Sb$_2$Se$_3$/In$_2$Se$_3$ remains a trivial semiconductor~\cite{li_Materials_Today_Physics_2026}.

Beyond time-reversal-symmetric scenerios, interface between a topological insulator and a ferromagnetic material gives rise to magnetic proximity effects. These effects break time-reversal symmetry and open a band gap in the Dirac-like edge or surface states, enabling phenomena such as the topological magnetoelectric effect~\cite{bhattacharyya_Advanced_materials_2021, grutter_PRM_2021, tokura_Nature_review_phy_2019magnetic, sharma_Nanotechnology_2025}. In addition, heterostructures formed by placing a two-dimensional topological insulator in close proximity to a conventional superconductor exhibit superconducting proximity effects, which can induce exotic quantum phases~\cite{Cook_PhysRevB_2011, qi_rev_modern_phy_2011, rachel_Reports_on_progress_in_phy_2018, wang_Nature_materials_2017}. Notably, such systems provide a platform for the emergence of Majorana bound states, quasiparticles with non-Abelian statistics, which hold significant promise for the realization of fault-tolerant topological quantum computing~\cite{fu_PRL_2008, kitaev_Physics_uspekhi_2001, lutchyn_PRL_2010, sarma_npj_quantum_information_2015, beenakker_Annual_rev_condensed_mat_phy_2013, leijnse_Semiconductor_science_and_technology_2012, fu_PRL_2008}.

 \begin{figure*}[hbtp]
\centering
\includegraphics [width=1\textwidth]{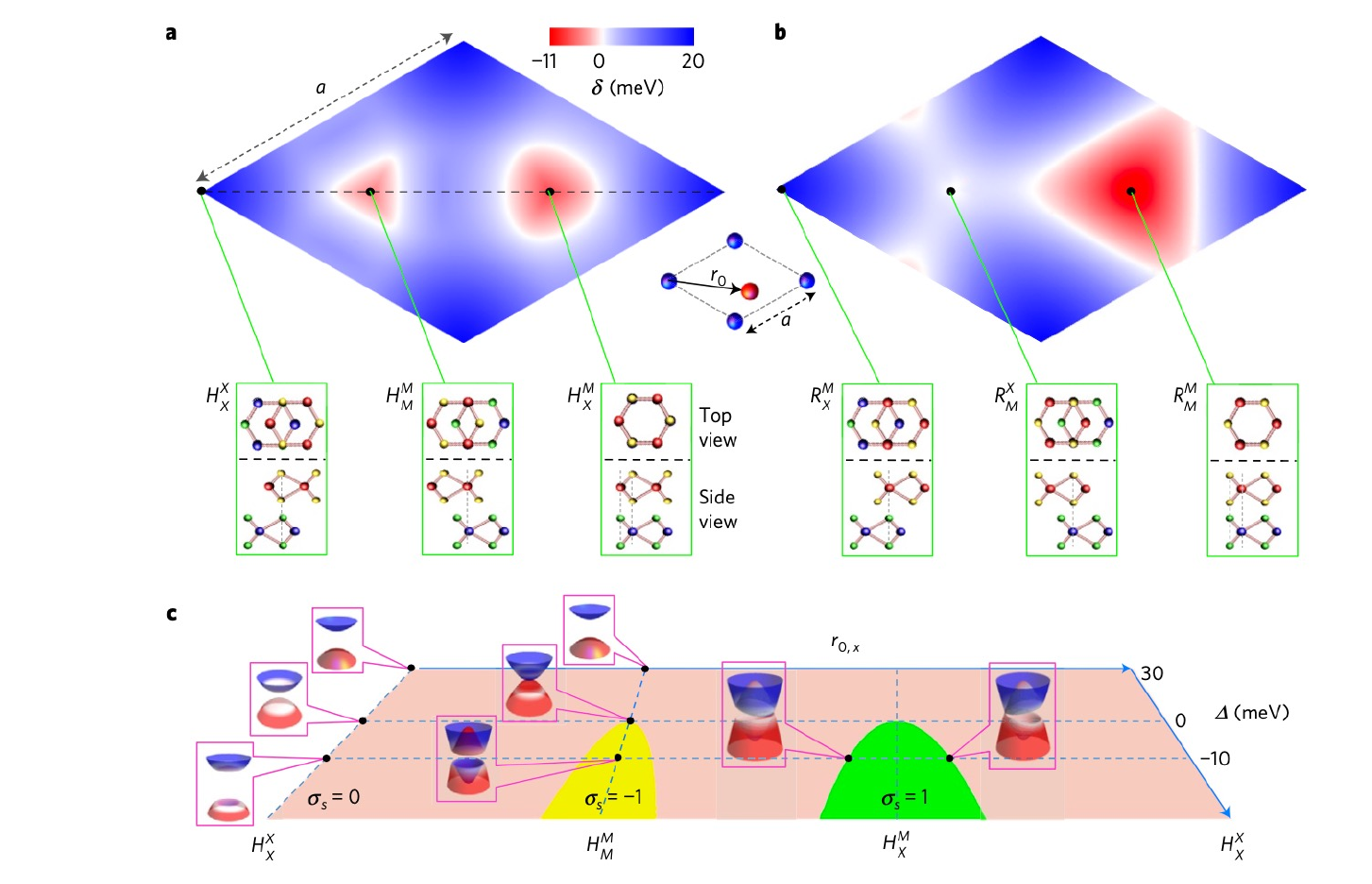}
\caption{Hybridization gap ($\delta$) as a function of the relative interlayer translation r$_0$ for (a)\,H-type stacking and (b)\,R-type stacking of TMD bilayers with inverted type-II band alignment ($\Delta$\,=\,-40 meV). Regions shown in red correspond to a topological insulating phase, while blue regions indicate a normal insulating phase. (c)\,Phase diagram as a function of both the interlayer band offset 
$\Delta$ and the translation vector r$_0$ for H-type stacking, with r$_0$ constrained along the dashed trajectory shown in panel (a). The yellow and green regions represent two distinct quantum spin Hall insulating phases, characterized by different spin Hall conductances within the hybridization gap. Insets show the hybridized Dirac dispersions at the K valley for selected representative points in the phase space (Figure reprinted with permission from Ref.~\cite{tong_Nature-phy_2017}).}
\label{fig:Twisting}
\end{figure*}

\subsection{Twisting}

Since the discovery of correlated insulating and superconducting phases in twisted bilayer graphene, moiré engineering has emerged as a powerful platform for realizing novel quantum phenomena in low-dimensional systems~\cite{cao_Nature_2018_correlated, cao_PRL_2016, bistritzer_PNAS_2011, suarez_PRB_2010, lopes_PRB_2012, kim_PNAS_2017, cao_Nature_2018_superconductivity, wu_PRL_2018}. Moiré materials are typically engineered either by introducing a relative twist angle between adjacent layers in vdW heterostructures or by exploiting lattice-mismatch–induced strain in heterobilayers~\cite{bistritzer_PNAS_2011, cao_Nature_2018_correlated, andrei_Nature_rev_materials_2021, kennes_Nature_Physics_2021, du_Science_2023, Dey2023, Dey2024, Dey2025}. Both approaches generate long-wavelength moiré interference patterns, leading to the formation of large superlattice unit cells with nanometer-scale periodicity. These emergent periodic potentials strongly renormalize the electronic band structure by suppressing the kinetic energy of otherwise dispersive bands, thereby enhancing the role of electron–electron Coulomb interactions. The resulting flat or nearly flat bands provide an ideal platform for correlated electron physics, giving rise to a diverse range of emergent quantum phases, including two-dimensional Mott insulating states~\cite{li_Nature_2021, regan_Nature_2020, po_PRX_2018} and Wigner crystal formation~\cite{regan_Nature_2020, matty_Nature_com_2022, padhi_PRB_2021, li_Nature_2021, nuckolls_Nature_Review_materials_2024}.

Importantly, moiré materials can also host topological phases in ways that are fundamentally distinct from the conventional single-particle picture~\cite{pan_PRL_2022, zhou_arxiv_2026, adak_Nature_review_materials_2024}. Moiré flat bands emerge either through interlayer hybridization or via Brillouin-zone folding of the low-energy dispersive bands of the constituent layers without requiring an external magnetic field. These effects produce narrow bandwidth electronic bands with an intrinsic Berry curvature inherited from the individual layers. Since these flat bands preserve time-reversal symmetry, applying an external magnetic field can unlock the hidden Berry curvature, stabilizing magnetic topological phases even in systems with negligible $d$-electron contributions or weak spin–orbit coupling, as in graphene-based moiré materials~\cite{nuckolls_Nature_Review_materials_2024, andrei_Nature_rev_materials_2021}. Consequently, moiré systems can stabilize novel correlated topological phases, including some with no single-particle analogue, such as fractional Chern insulators (FCIs)~\cite{li_PRX_2021, wang_PRL_2024, xie_Nature_2021, ledwith_Phy_rev_reserach_2020}.

Extensive studies have focused on TRS-broken topological phases in moiré supercells. For instance, twisted bilayer graphene (TBG) has been explored for FCIs at various fillings~\cite{xie_Nature_2021, ledwith_Phy_rev_reserach_2020, liu_arxiv_2022}, while transition-metal dichalcogenides (TMDs) have been investigated for realizing fractional quantum anomalous Hall phases in twisted bilayers~\cite{reddy_PRB_2023, zhao_acs_nano_2025, cai_Nature_2023, xu_PRX_2023}. Several reviews provide detailed discussions of these magnetic topological phases~\cite{morales_Nature_rev_phy_2024, nuckolls_Nature_Review_materials_2024, kennes_Nature_Physics_2021, adak_Nature_review_materials_2024}. However, in this Research Update, we focus instead on TRS-preserving topological phases and the emerging field of moiré supercells with locally varying topological character.

Conventional QSH insulators exhibit an insulating bulk and topologically protected helical edge states composed of counterpropagating modes with opposite spin polarization, resulting in a quantized edge conductance of $G_0=e^2/h$~\cite{hasan_Review_of_modern_phy_2010, Murakami_Science_2003, qi_rev_modern_phy_2011}. While most experimentally realized QSH systems display integer quantization, Kang \textit{et al.}~\cite{kang_Nature_2024} reported the emergence of a fractional QSH phase in 2.1$\degree$-twisted MoTe$_2$. At filling factor $\nu$\,=\,3, the edge states contribute a fractional conductance of $3/2G_0$, accompanied by a vanishing anomalous Hall conductivity, hence preserving the time-reversal symmetry. Furthermore, twisted bilayer WSe$_2$~\cite{kang_Nano_Letters_2024} hosts QSH phases with one and two pairs of helical edge states at moiré hole filling factors $\nu$\,=\,2 and $\nu$\,=\,4, respectively.

First-principles calculations and symmetry analysis identify the BaCu monolayer as a two-dimensional topological insulator~\cite{liang_npj_computational_materials_2025}. When combined with a 30$\degree$ twisted h-BN supercell, the resulting $\alpha$/$\beta$-BaCu/BN heterobilayers host 2D Weyl and type-III Dirac points, demonstrating twist-angle-driven topological modulation~\cite{liang_npj_computational_materials_2025}. Furthermore, ZrS$_2$ heterostructures at small twist angles can host a quantum spin Hall insulating phase, where the emergence of this phase is driven by the interplay of geometric frustration and strong spin-orbit coupling~\cite{claassen_Nature_comm_2022}.

A recently emerging research direction aims to engineer spatially programmable topological phases within a single moiré supercell, allowing the coexistence of topologically trivial and nontrivial regions in the same material. Using $k$ $\cdot$ $p$ perturbation theory and tight-binding Hamiltonians, Tong \textit{et al}.~\cite{tong_Nature-phy_2017} demonstrated that moiré superlattices formed by twisted heterobilayers of massive Dirac materials, such as transition-metal dichalcogenides, exhibit spatially varying local stacking configurations. This variation gives rise to alternating topological insulator and normal insulator domains within a single moiré supercell, thereby forming a topological mosaic, as illustrated in Fig.~\ref{fig:Twisting}. This behavior is primarily governed by the spatial modulation of interlayer hybridization between massive Dirac cones, which depends sensitively on the local atomic registry across the moiré supercell. Furthermore, strain provides an additional degree of freedom to continuously tune both the topology and geometry of the mosaic, enabling a transition from two-dimensional arrays of TI nanodots to one-dimensional arrays of TI nanostripes.

In a similar context, twisted bilayer of Bi$_2$(Te$_{1-x}$Se$_x$)$_3$~\cite{tateishi_PRR_2022} has been proposed to host coexisting topological insulator and normal insulator domains, with domain interfaces supporting edge states that form nearly flat bands and dominate the low-energy electronic properties~\cite{tateishi_PRR_2022}. These domain-boundary states further drive a moiré-scale band inversion, giving rise to emergent moiré topological phases with associated moiré-scale edge states\,--\,an “edge state from edge state” hierarchy unique to twisted bilayer Bi$_2$(Te$_{1-x}$Se$_x$)$_3$. Further, Wang \textit{et al}.~\cite{wang_PRB_2023} investigated the role of these periodically arranged edge states at the interfaces between trivial and nontrivial domains in determining the global topological properties. By developing a continuum model that incorporates C$_{6z}$ and 
C$_{3z}$ rotational symmetries, they demonstrated that the global topological phase transition at the charge neutrality point (CNP) can be controlled by the size of the domain walls and the moiré period. However, the role of structural relaxation on the topological character of local domains remains unexplored in these studies.


\begin{figure*}[hbtp]
\centering
\includegraphics [width=16cm]{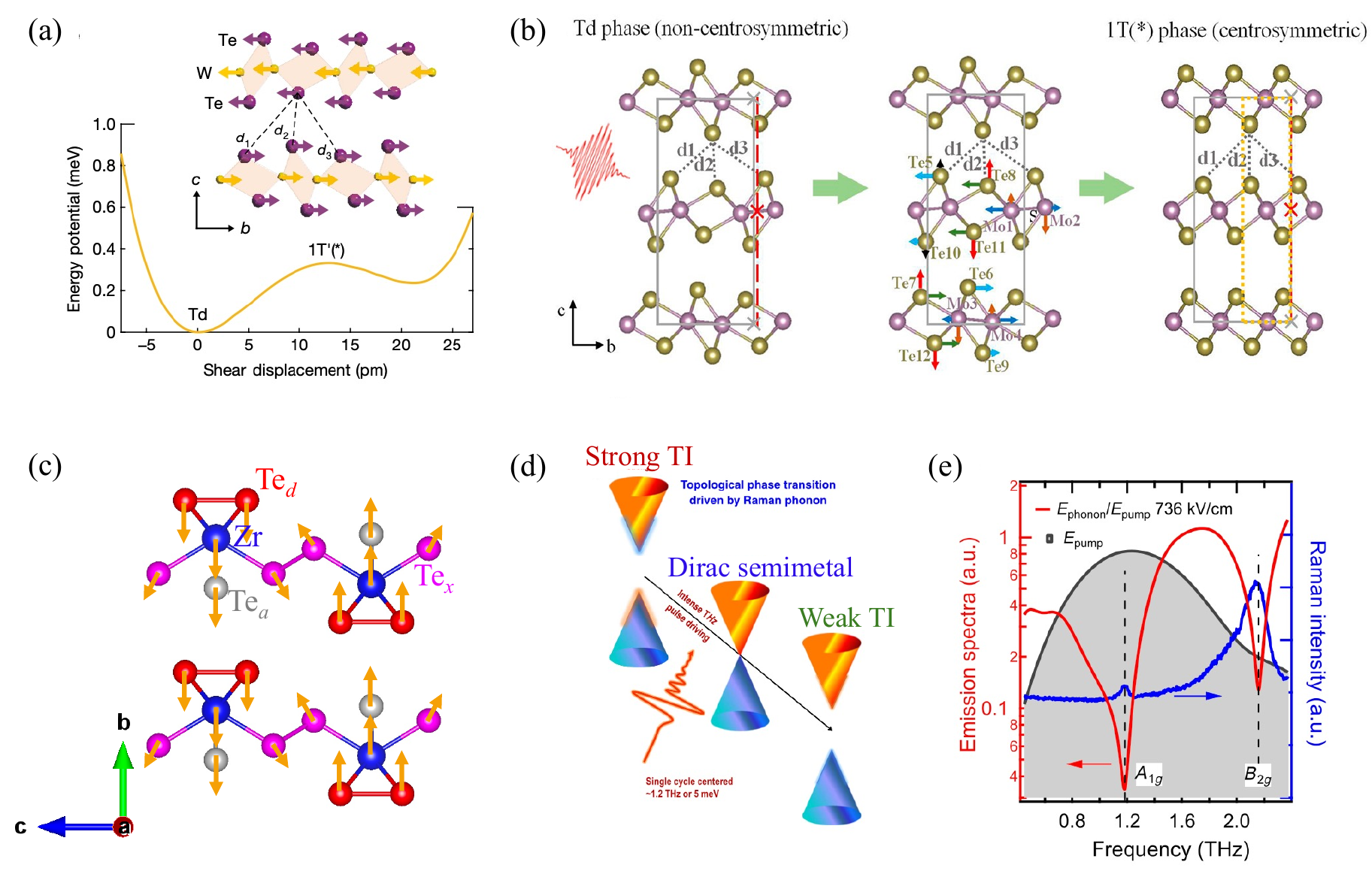}
\caption{Light–matter interaction–driven topological phase transitions. Top panel: (a)\,Potential energy as a function of interlayer shear displacement in WTe$_2$, illustrating the shear-induced transition from the non-centrosymmetric T$_d$ phase to the centrosymmetric 1T'(*) phase~\cite{sie_Nature_2019}. (b)\,THz-light–induced intra- and interlayer structural transitions in T$_d$-MoTe$_2$~\cite{qi_ACS_nano_2022}. Bottom panel: Photoinduced phase transition in ZrTe$_2$. (c)\,Atomic displacement pattern of the A$_{1g}$ phonon mode. (d)\,Schematic of the transition from a strong topological insulator to a weak topological insulator via a Dirac semimetal. (e)\,Coherent phonon spectrum (red) extracted from time-domain traces at the maximum THz field strength, overlaid with the pump THz spectrum (gray)~\cite{vaswani_PRX_2020}}
\label{fig:light-matter}
\end{figure*}

\subsection{Light-matter interaction}
Another highly versatile platform for achieving dynamical control over nonequilibrium topological phase transitions is light–matter interaction. Recent experimental and theoretical advances have demonstrated that electromagnetic radiation can strongly couple to various degrees of freedom in quantum materials, including lattice vibrations, electronic charge, spin, and orbital degrees of freedom~\cite{bao_Nature_Rev_Phy_2022,yang_Nature_rev_materials_2023, truc_PNAS_2024, wang_physics_status_solidi_RRL_2013}. Such couplings can profoundly reshape the electronic structure and give rise to emergent nonequilibrium quantum phases.

Compared to conventional approaches for tuning topological phases, such as the application of strain, static electric fields, or chemical doping, which often require direct physical contact via electrodes, tips, or wiring, light–matter interaction offers a fundamentally contactless and noninvasive means of control~\cite{disa_Nature_Physics_2021, bao_Nature_Rev_Phy_2022, yang_Nature_rev_materials_2023}. This intrinsic advantage makes optical manipulation particularly attractive for ultrafast, reversible, and spatially selective control of different properties in quantum materials.

In the context of topological phase transitions, light can induce topological states through various mechanisms. One prominent route is Floquet engineering~\cite{lindner_Nature_Phy_2011, katan_PRL_2013, narayan_PRB_2015, roy_PRB_2017, liu_PRL_2018, bao_Nature_Rev_Phy_2022, kumar_Communication_Physics_2022}, in which light acts as a time-periodic electric field that renormalizes the electronic band structure. Alternatively, resonant light–matter coupling in the terahertz (THz) regime can directly interact with electronic and lattice degrees of freedom, leading to a substantial modification of the energy landscape in the excited state~\cite{zhang_PRX_2019, sie_Nature_2019, yang_Nature_rev_materials_2023, wei_arxiv_2025}. In this scenario, the energy carried by incident photons is redistributed among different degrees of freedom through nonlinear couplings, driving the system into a nonequilibrium excited state. Such processes can stabilize hidden or metastable states that host structural and topological phases distinct from the equilibrium ground state, thereby enabling topological phase transitions beyond the Floquet picture~\cite{forst_Nature_PhysicS-2011, disa_Nature_Physics_2021, mankowsky_Report_on_progress_in_phy_2016, guan_PRL_2022, zeiger_PRB_1992, bartels_Chemical_phy_letters_2003, li_Science_2019}.

In vdW materials, the latter approach is particularly compelling, as their electronic properties and consequently their topological character are strongly governed by stacking configuration and interlayer coupling~\cite{fox_Chemical_reviews_2023, bordoloi_2D_Materials_2025, kou_APL_2018, bao_Nature_Rev_Phy_2022}. Light–matter interaction can significantly modify these factors by selectively exciting low-energy interlayer phonon modes, such as shear and breathing modes, which alter relative stacking, interlayer distances, and crystal symmetry~\cite{qi_ACS_nano_2022, zhang_PRX_2019, sie_Nature_2019, qi_ACS_nano_2022, Singh_PRL_2020, vaswani_PRX_2020, bao_Nature_Rev_Phy_2022}. As a result, optical excitation provides a powerful route to dynamically engineer interlayer interactions and access nonequilibrium topological phases unique to vdW heterostructures.

Transition metal dichalcogenides (TMDs) have been extensively studied to investigate the effect of THz light pulses on topological phase transitions. In particular, MoTe$_2$~\cite{qi_ACS_nano_2022, zhang_PRX_2019} and WTe$_2$~\cite{sie_Nature_2019, qi_ACS_nano_2022} have been observed to undergo intriguing topological transitions upon excitation of interlayer shear-phonon modes by light. Weyl semimetal WTe$_2$, THz pulses can generate large interlayer shear strains [Fig.~\ref{fig:light-matter}(a)], producing a metastable phase with a distinct topological character. Nonlinear optical studies reveal that this transition involves a structural transformation from the noncentrosymmetric T$_d$ phase to the topologically trivial centrosymmetric 1T$^{\prime}$(*) phase~\cite{sie_Nature_2019}. Further, Qi \textit{et al.}~\cite{qi_ACS_nano_2022} showed that photoexcitation in XTe$_2$ (X = Mo, W) triggers simultaneous interlayer and intralayer structural changes as shown in Fig.~\ref{fig:light-matter}(b), where stretching of Mo–Mo bonds and reduction of out-of-plane Te wrinkling rapidly suppress Peierls distortion, inducing a transition to an intralayer 1T-like structure~\cite{qi_ACS_nano_2022}.

Vaswani \textit{et al.} demonstrated that by coherently exciting the lowest experimentally observed Raman-active mode in ZrTe$_5$, which has A$_{1g}$ symmetry, the system can be transitioned from a strong TI to a Dirac semimetal, and subsequently to a weak TI (Fig.~\ref{fig:light-matter}(c)-(e))~\cite{vaswani_PRX_2020}. This symmetric eigenmode, which corresponds to opposite displacements of Te atoms within the same layer [Fig.~\ref{fig:light-matter}(c)], modulates the vdW interlayer coupling and, in turn, controls the band inversion mechanism at the $\Gamma$ point. Consequently, coherent excitation of this phonon drives a topological phase transition without breaking the underlying crystal symmetry. On the other hand, Luo \textit{et al.}~\cite{luo_Nature_Materials_2021} observed the dynamical photogeneration of a pair of Weyl points in ZrTe$_5$ via laser-induced coherent excitation of the B$_{1u}$ mode, the lowest infrared-active phonon, which breaks inversion symmetry. The chirality of these Weyl points manifests as a transverse, helicity-dependent photocurrent, perpendicular to the axis of dynamical inversion symmetry breaking, through the circular photogalvanic effect~\cite{luo_Nature_Materials_2021}.

Additionally, trigonal bismuthene exhibits a light-induced transition from a quantum spin Hall state to a quantum anomalous Hall state with Chern number ±3, followed by a subsequent transition to a compensated Chern-insulating state ($C$ = 0)~\cite{li_arxiv_2025}. Photoexcitation has been shown to simultaneously drive structural and topological phase transitions in ferroelectric and paraelectric bismuth monolayers~\cite{Peng_PhysRevLett_2024}. By tuning the photoexcited carrier density and electronic temperature, four previously hidden crystal structures emerge, two of which realize distinct quantum spin Hall insulating phases characterized by different spin Chern numbers. In addition, light-induced transitions transform the ferroelectric and paraelectric phases into two-dimensional Dirac and nodal-line semimetals, respectively, each hosting characteristic topological edge states~\cite{Peng_PhysRevLett_2024}. 

On the other hand, a different mechanism for light-induced topological phase transitions, driven purely by electronic effects without involving the lattice, has been demonstrated in TaAs~\cite{sirica_Nature_materials_2022}. Femtosecond optical excitation transiently lowers the magnetic point symmetry of this type-I Weyl semimetal on a picosecond timescale, enabling Weyl node manipulation and showing that light can dynamically control electronic symmetries and thus topological properties on ultrafast timescales.

\subsection{Chemical functionalization}

Another widely investigated route to induce and tailor topological phase transitions in vdW materials is chemical or surface functionalization via adatom adsorption. Adsorbed adatoms can substantially modify the electronic structure of pristine 2D systems and drive topological phase transitions through several distinct mechanisms~\cite{zhang_Advanced_functional_materials_2021, Huertas_PhysRevB_2006, Min_PhysRevB_2006}. Some key mechanisms include enhancement of the effective SOC strength via hybridization with heavy adatoms, tuning of the chemical potential through charge transfer, and breaking of certain symmetries, such as inversion or time-reversal symmetry, in the resulting structure~\cite{Brey_PhysRevB_2015, Qiao_PhysRevB_2010, Weeks_PhysRevX._2011, Acosta_PhysRevB_2014}. Symmetry breaking can give rise to Rashba spin–orbit coupling or magnetic exchange interactions, depending on the nature of the adatom. Collectively, these effects can either stabilize entirely new topological phases in the resulting materials or enhance the band gap of systems that are already 2D TIs, thereby improving their prospects for room-temperature operation.

Graphene, the most extensively studied 2D material, has served as a prototypical platform for exploring adatom-induced topological engineering. In principle, pristine graphene is a quantum spin Hall insulator~\cite{Hasan_RevModPhys_2010, Qi_RevModPhys_2011}; however, owing to its extremely weak intrinsic SOC, on the order of 25–50 $\mu$eV~\cite{Huertas_PhysRevB_2006, Min_PhysRevB_2006}, the resulting topological band gap is vanishingly small. This renders pristine graphene unsuitable for room-temperature 2D TI applications. Numerous theoretical and experimental studies have shown that selective adsorption of adatoms can substantially enhance the SOC in graphene, thereby opening a much larger bulk band gap. 

\begin{figure}
\centering
\includegraphics [width=8cm]{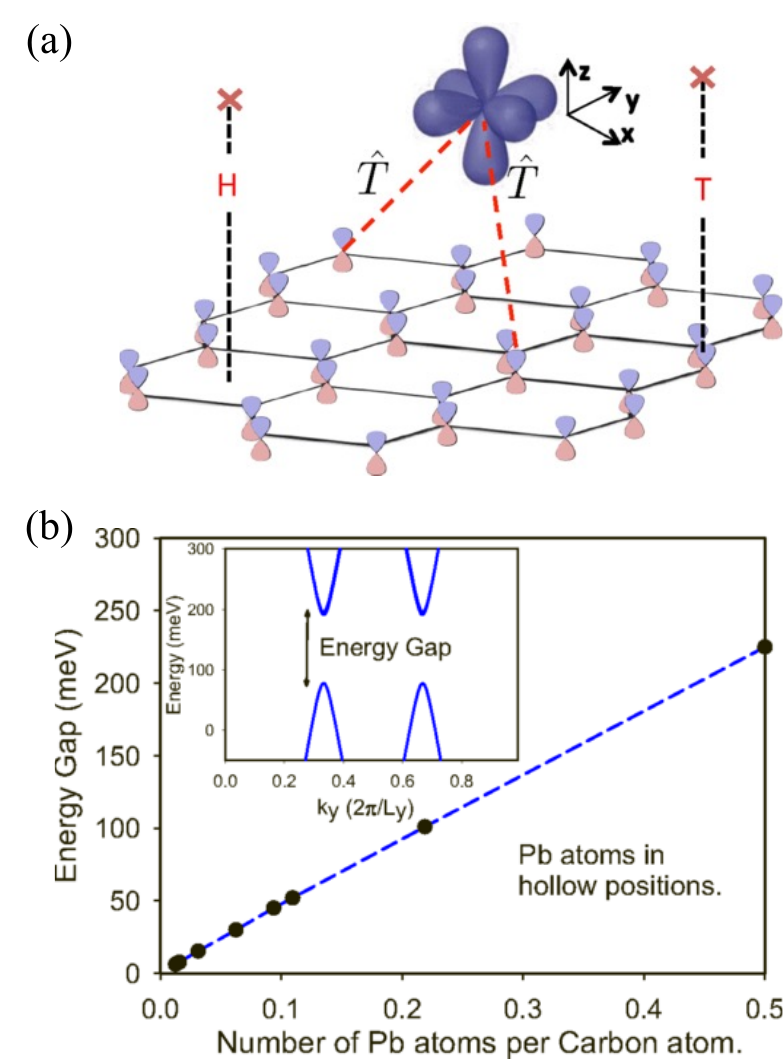}
\caption{(a) Schematic illustration of an adatom with outer p electrons inducing effective hopping between carbon atoms in graphene. H and T denote the hollow and top adsorption sites, respectively. (b) Variation of the energy gap opened at the Dirac point by Pb adatoms. The inset shows the band structure of graphene in a rectangular supercell with a Pb coverage of 0.25 Pb atoms per carbon (Figure reprinted with permission from Ref.~\cite{Brey_PhysRevB_2015}).}
\label{fig:adatom}
\end{figure}

\begin{figure*}[hbtp]
\centering
\includegraphics [width=1\textwidth]{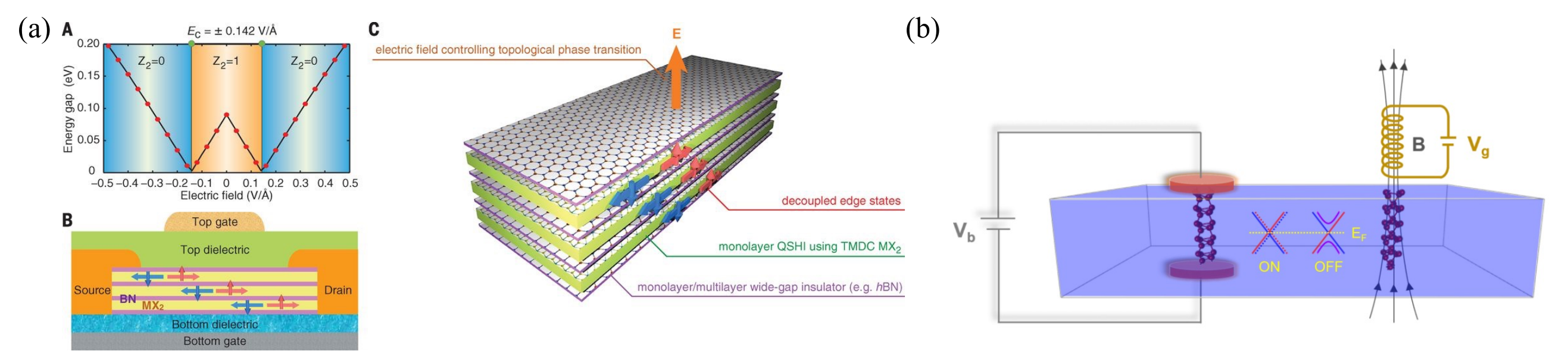}
\caption{Panel (a) illustrates the proposed TFET architecture based on vdW heterostructures composed of 2D transition-metal dichalcogenides MX$_2$(M = Mo, W; X = S, Se, Te)  and the wide-bandgap insulator h-BN (Figure adapted from Ref.~\cite{qian_Science_2014}). (b)\,The theoretically proposed high-fidelity TFET model that exhibits quantized conductance in both the ON and OFF states (Figure reprinted with permission from Ref.~\cite{li_PRB_2023}).}
\label{fig:TFET}
\end{figure*}

In particular, adsorption of heavy $p$-block elements, such as Pb~\cite{Brey_PhysRevB_2015}, has been shown to induce SOC strengths of $\sim$ 40meV at adatom concentrations of about 0.1 per carbon atom. Microscopically, the enhancement arises from hybridization of graphene with the $p$ orbitals of the adatoms, which introduces additional hopping channels between carbon atoms (Figure~\ref{fig:adatom}). For adatoms adsorbed at hollow sites, spin-conserving tunneling occurs, leading to an enhancement of intrinsic SOC. In contrast, a random distribution of adatoms induces spin-flip hopping processes, giving rise to Rashba-type SOC~\cite{Brey_PhysRevB_2015}.
For other heavy adatoms, such as indium and thallium, even modest coverages (6$\%$) are predicted to generate significant band gaps, equivalent to $\sim$80 K and $\sim$240 K, respectively~\cite{Weeks_PhysRevX._2011}. Furthermore, codoping graphene with thallium and tetrafluorotetracyanoquinodimethane (F4-TCNQ) has been reported to open a band gap of 26.9\,meV~\cite{Yang_JPDP_2016}.

Additionally, experimental studies have demonstrated that graphene grown on Cu substrates exhibits a SOC–induced band splitting of approximately 20 meV~\cite{balakrishnan_Nature_comm_2014}. Marchenko \textit{et al}. reported a giant enhancement of SOC splitting, reaching values of about 100 meV, with Au intercalation at the graphene–Ni interface\cite{marchenko_Nat_Comm_2012}. This pronounced increase has been attributed to strong hybridization with Au-$5d$ orbitals, as revealed by angle-resolved photoelectron spectroscopy measurements. Notably, fluorinated graphene has been shown to exhibit an induced SOC strength that is nearly three orders of magnitude larger than that of pristine graphene, and approximately an order of magnitude larger than that of hydrogenated graphene~\cite{Irmer_PhysRevB_2015}. Moreover, Os, Ir, Cu–Os dimers, and Cu–Ir dimers have been shown to significantly enhance the bandgap in graphene~\cite{Hu_PhysRevLett_2012}.

Beyond time-reversal-invariant topological phases, Qiao \textit{et al.} demonstrated that the QAH effect can be realized in graphene by simultaneously introducing Rashba SOC and an exchange field highlighting the potential of adatom and interface engineering for stabilizing magnetic topological phases in graphene~\cite{Qiao_PhysRevB_2010}. Ru adatoms can also stabilize different topological phases, including QSH, QAH, and trivial insulating phases, depending on their concentration~\cite{Acosta_PhysRevB_2014}. 

Beyond graphene, silicene has also been shown to exhibit tunable topological behavior through doping with $4d$ transition metal atoms~\cite{Zhang_PhysRevB_2013}. The interplay between transition metal induced exchange interactions, SOC, and the staggered AB sublattice potential gives rise to a rich variety of topological phases, including QAH states, valley Hall states, and valley-polarized metallic states. Notably, QAH states with different Chern numbers can be realized depending on the choice of dopant: for instance, a Chern number of -2 is obtained in Nb or Ru doped silicene, whereas Y doped silicene exhibits a Chern number of +1~\cite{Zhang_PhysRevB_2013}. 
Similarly, germanene, a 2D TI with a band gap of 23.4 meV in its freestanding monolayer form, can achieve a significantly enhanced gap of 0.3 eV upon passivation with iodine atoms~\cite{Xu_PhysRevLett_2013}. Arsenene, when functionalized as AsX (X = F, OH, or CH$_3$), behaves as a 2D TI with a band gap in the range of 0.1-0.16 eV~\cite{zhao_Nanoscale_2016}. While pristine TlP is topologically trivial, functionalization with halogen atoms can induce a topological phase transition~\cite{Chen_RSC_advances_2023}, highlighting the general utility of chemical functionalization for engineering topological properties in two-dimensional material.

\section{Emerging application oriented aspects}

The gradual slowdown of Moore’s law has led to a higher demand for device miniaturization and has renewed research efforts to identify viable alternatives to conventional semiconductor-based electronic technologies~\cite{moore_Electronics_magazine_1965, mack_IEEE_transactions_on_semiconductor_manufcaturing_2011, schaller_IEEE_Spectrum_2002}. For instance, spintronic devices have emerged as promising alternatives, as they exploit the electron’s spin degree of freedom rather than its charge, offering potential advantages in energy efficiency and enhanced functionality.

In a similar context, 2D TIs have attracted considerable attention as next-generation device platforms, primarily due to the robustness of their topologically protected edge states against nonmagnetic disorder and defects, along with the high conductivity of these edge channels~\cite{Murakami_Science_2003, Chang_Science_2013, Hasan_RevModPhys_2010, weber_Journal_of_phy_mat_2024}. These unique characteristics make 2D TIs particularly appealing for low-power electronic and spintronic applications. In this section, we briefly summarize some of the key application prospects of 2D TIs. While several comprehensive review articles provide detailed discussions of these applications~\cite{tian_Materials_2017, gilbert_Communication_Phy_2021, jin_Nanoscale_2023, zhang_Small_2022}, our focus here is to present a concise update on recent progress in application-oriented research. The associated challenges and future opportunities are discussed in the subsequent section.

\begin{description}[leftmargin=0cm, labelwidth=5cm, labelsep=0.5em, align=left, style=nextline]
\item[Topological field effect transistors (TFETs)]

One of the most common applications of 2D TIs is in TFETs. Conventional field-effect transistors are fundamental building blocks of modern semiconductor electronics, where the ON/OFF mechanism relies on switching between high- and low-conductivity states~\cite{dacey_Bell_system_technical_journal-1955, cheng_Nature_electronics_2022}. A large ON/OFF ratio is essential for achieving high device performance. However, in the semiconductor materials used in conventional FETs, charge carriers undergo significant scattering, resulting in substantial energy dissipation in the form of heat~\cite{freitag_Nano_Letters_2009, pop_Proceedings_IEEE_2006, alam_APL_2019}.
TFETs address this limitation by exploiting topologically protected edge states, which act as robust, dissipationless transport channels. These edge states can be viewed as a two-lane highway for electrons with opposite spins, enabling spin-momentum–locked transport with strongly suppressed backscattering. In most reported TFET architectures, the ON state is governed by conducting topological edge states, while the OFF state is achieved by driving a transition to a topologically trivial insulating phase~\cite{gilbert_Communication_Phy_2021, jin_Nanoscale_2023, zhang_Small_2022}.

Initial efforts to design TFETs have primarily focused on the well known topological insulator Bi$_2$Se$_3$, employing it as the channel in dual-gate FET architectures~\cite{liu_APL_2011, steinberg_Nano_Letter_2010}. However, unintentional selenium vacancies in Bi$_2$Se$_3$ lead to high bulk carrier densities, creating low-resistance conduction paths that suppress the contribution of topological surface states to overall conductivity~\cite{hor_PRB_2009, steinberg_Nano_Letter_2010, cho_Nano_Letters_2011}. To mitigate this parallel conduction through bulk and surface channels, ultrathin Bi$_2$Se$_3$ films were used, enhancing quantum confinement and reducing bulk state contributions~\cite{cho_Nano_Letters_2011}. By decreasing the film thickness, bulk conductivity is suppressed, allowing electrostatic gating to effectively deplete bulk carriers and achieve a true OFF state. Nonetheless, thickness reduction alone is insufficient to induce a transition from $n$-type to $p$-type dominant doping. This can be realized by calcium doping of Bi$_2$Se$_3$~\cite{checkelsky_PRL_2011}, which enables surface-state-dominated conduction up to ~130 K. A significant limitation arises when reducing the thickness below ~6 nm, a hybridization gap opens between the top and bottom Dirac surface states, giving mass to the Dirac fermions and restricting further thinning~\cite{taskin_PRL_2012}.

On the other hand, Qian \textit{et al.}~\cite{qian_Science_2014} proposed a TFET based on vdW heterostructures composed of 2D transition metal dichalcogenide MX$_2$(M = Mo, W; X = S, Se, Te) quantum spin Hall layers separated by a wide–bandgap insulator h-BN [Figure~\ref{fig:TFET}(a)]. The h-BN spacer suppresses interlayer coupling between the edge states of adjacent QSH layers, thereby preventing the opening of a hybridization-induced band gap. In the absence of external perturbations, the system remains in a topologically nontrivial state with Z$_2$=1, corresponding to the ON state of the transistor~\cite{qian_Science_2014}. Upon applying a moderate electric field, however, the 1T'-phase MX$_2$ undergoes a topological phase transition to a trivial Z$_2$=0 state, effectively switching the TFET to the OFF state. On the other hand, Na$_3$Bi~\cite{collins_Nature_2018} and NaIrTe$_2$~\cite{Li_PhysRevB_2020} have also been reported to exhibit electrically tunable topological phase transitions, highlighting their potential suitability for TFET applications.

However, most reported TFETs still suffer from nonzero leakage current in the OFF state, similar to many conventional FETs. To address this issue, Li \textit{et al.}~\cite{li_PRB_2023} proposed a proof-of-concept TFET [Figure~\ref{fig:TFET}(b)] that exhibits quantized conductance in both the ON and OFF states, enabled by helical and chiral topological screw dislocation states in three-dimensional TIs~\cite{li_PRB_2023}. In this design, one screw dislocation serves as the conduction channel, while the other functions as a gate whose transport properties are controlled by an external magnetic field. Using BaBiO$_3$ as a prototypical material, they demonstrated an ON-state conductance of $2e^2/h$ and an OFF-state conductance of $e^2/h$. This strategy offers a promising route toward high-fidelity TFETs.

\item[Topological memory devices] 

2D TIs can also be used to design memory devices, including magnetic random-access memory (MRAM)~\cite{wu_Nature_comm_2021}, topological switching random-access memory (TRAM)~\cite{zhang_Small_2022}, and resistive random-access memory (RRAM)~\cite{zhang_Nature_materials_2019, bao_JPDAP_2018, bryja_2D_Materials_2021}. 2D TIs are well suited for these applications due to their strong spin–orbit coupling, large spin–orbit torque, and topological phase transitions that can be tuned by strain and electric fields. Beyond high operating speed and low energy consumption, SOT-MRAM devices based on 2D vdW TI heterostructures offer faster read and write operations. This enhanced performance arises from the strong spin–orbit torque exerted on adjacent magnetic layers by the spin-momentum–locked topological surface or edge states. On the other hand, materials such as Bi$_2$Te$_3$~\cite{bao_JPDAP_2018}, Bi$_2$Se$_3$~\cite{das_JAP_2018}, Sb$_2$Te$_3$~\cite{bryja_2D_Materials_2021} and MoTe$_2$ are widely explored for RRAM applications, while GeTe/Sb$_2$Te$_3$ superlattices have shown strong potential for TRAM devices~\cite{zhang_Small_2022}.

\item[Topological p-n junctions]

Another application of topological materials is the topological p–n junction~\cite{zhang_Small_2022, gilbert_Communication_Phy_2021, jin_Nanoscale_2023}. In conventional semiconductors, a p–n junction is formed at the interface between p-type and n-type regions, allowing unidirectional current flow and serving as a key building block in diodes, transistors, and integrated circuits. Topological p–n junctions, however, do not operate in the same way as conventional p–n junctions~\cite{frisenda_Chemical_Society_reviews_2018, gilbert_Communication_Phy_2021}. They are formed by p- and n-doped topological materials, and at the interface, a standing wave arises due to interference between incident and reflected electronic waves~\cite{wang_PRB_2011}. Furthermore, in the presence of an external magnetic field, a gapless chiral edge state emerges, whose properties can be tuned by the applied gate voltage and magnetic field~\cite{Wang_PhysRevB_2012}.

\end{description}

Beyond the applications discussed above, 2D TIs also find use in a wide range of technologies, including spintronic devices~\cite{he_Nature_materials_2022, jin_Nanoscale_2023}, topological magnetic devices~\cite{zhang_Advanced_functional_materials_2025, chen_Newton_2026}, thermoelectric devices~\cite{xu_npj_quantum_materials_2017, ivanov_physica_status_solidi_b_2018, skinner_Reports_on_Progress_in_Phuysics_2026}, 
functional optoelectronic devices~\cite{Chorsi2022, Breunig2022},
and photodetectors~\cite{wang_Journal_of_Materials_chem_C_2020, bao_Journal_of_Semiconductors_2025, yang_Materials_today_communication_2022}.

\section{Summary and outlook}

This Research Update summarizes recent advancements in the engineering of tunable two-dimensional topological phases. Since the discovery of nontrivial topology in graphene, significant research over the past two decades has focused on the study of TIs. Early efforts mostly have focused on identifying materials with intrinsic topological properties, where strong SOC drives band inversion. However, discovering materials with robust and tunable topological character remains a significant challenge. To address this, a growing research direction has emerged, aiming to either induce nontrivial topology in otherwise trivial semiconductors or drive transitions between distinct topological phases using external stimuli. In this article, we primarily cover four key approaches: stacking, moiré engineering, chemical functionalization, and light–matter interaction. 

Many of these regimes, such as light-induced phase transitions, stacking-induced emergence of nontrivial topology, and moiré-engineered effects, are still in their infancy and face considerable challenges that needs to be addressed. Despite these challenges, the field holds significant promise, offering pathways toward functional spintronic and topological devices. Below, we outline our perspective on the major challenges and highlight exciting opportunities for future research in tunable 2D topological systems.

\subsection*{Major challenges}

\begin{description}[leftmargin=0cm, labelwidth=5cm, labelsep=0.5em, align=left, style=nextline]

\item[Experimental challenges] 
While theoretical and computational studies of 2D TIs have advanced considerably, experimental realization remains in its early stages. Most known 2D TIs exhibit very small band gaps, making them highly sensitive to thermal fluctuations. Synthesizing materials that retain topological properties at room temperature, while remaining stable against perturbations and preserving desirable electronic characteristics, remains a significant challenge~\cite{kou_Journal_of_phy_chem_c_2017, tian_Materials_2017, muchler_Angewandte_Chemie_International_Edition2012}. Additionally, unintended vacancies or interstitial defects often introduce bulk carriers, which suppresses topological edge transport~\cite{shan_PRB_2011, hor_PRB_2009, steinberg_Nano_Letter_2010, cho_Nano_Letters_2011}. In moiré-engineered systems, achieving and maintaining the precise stacking angles required for reproducible topological behavior is particularly difficult, further complicating experimental implementation~\cite{shen_Int_journal_of_extreme_manufacturing_2026, lau_Nature_2022, zhang_ACS_applied_materials_2024}.

\item[Theoretical and computational Challenges]
Despite significant progress in theoretical modeling and computational studies of 2D TIs, several challenges remain. First, identifying suitable material candidates for vdW stacking to realize 2D TIs with desired properties is still difficult. Furthermore, accurate prediction of tunable topological phases under realistic conditions, including defects, disorder, and electron correlation effects, remains a major challenge. In addition, simulating topological properties in moiré heterostructures at atomic resolution is computationally demanding, limiting the ability to study large-scale or highly complex systems.

\item[Challenges in application-oriented development] 
Although topological quantum devices are considered a promising alternative to conventional semiconductor technologies, the field remains in its early stages. For applications such as TFETs, materials must possess a suitable band gap at room temperature and topological properties that can be reliably tuned via external stimuli, such as gate voltage or strain. Identifying such materials is challenging, and even when candidate materials exist, achieving scalable fabrication remains difficult.

Additional challenges arise in the ultrathin limit, where finite-size effects can induce interactions between edge states~\cite{taskin_PRL_2012}, opening a gap and potentially converting the material into a trivial insulator, thereby restricting the available material landscape. Furthermore, precise tuning of carrier density, suppression of bulk conductivity while maintaining robust surface or edge transport, and switching the dominant carrier type between $n$- and $p$-type remain significant obstacles for practical device implementation. Addressing these issues requires further experimental and theoretical efforts to enable application-oriented development of 2D TIs.

\end{description}

\begin{figure}[!!t]
\centering
\includegraphics [width=1\columnwidth]{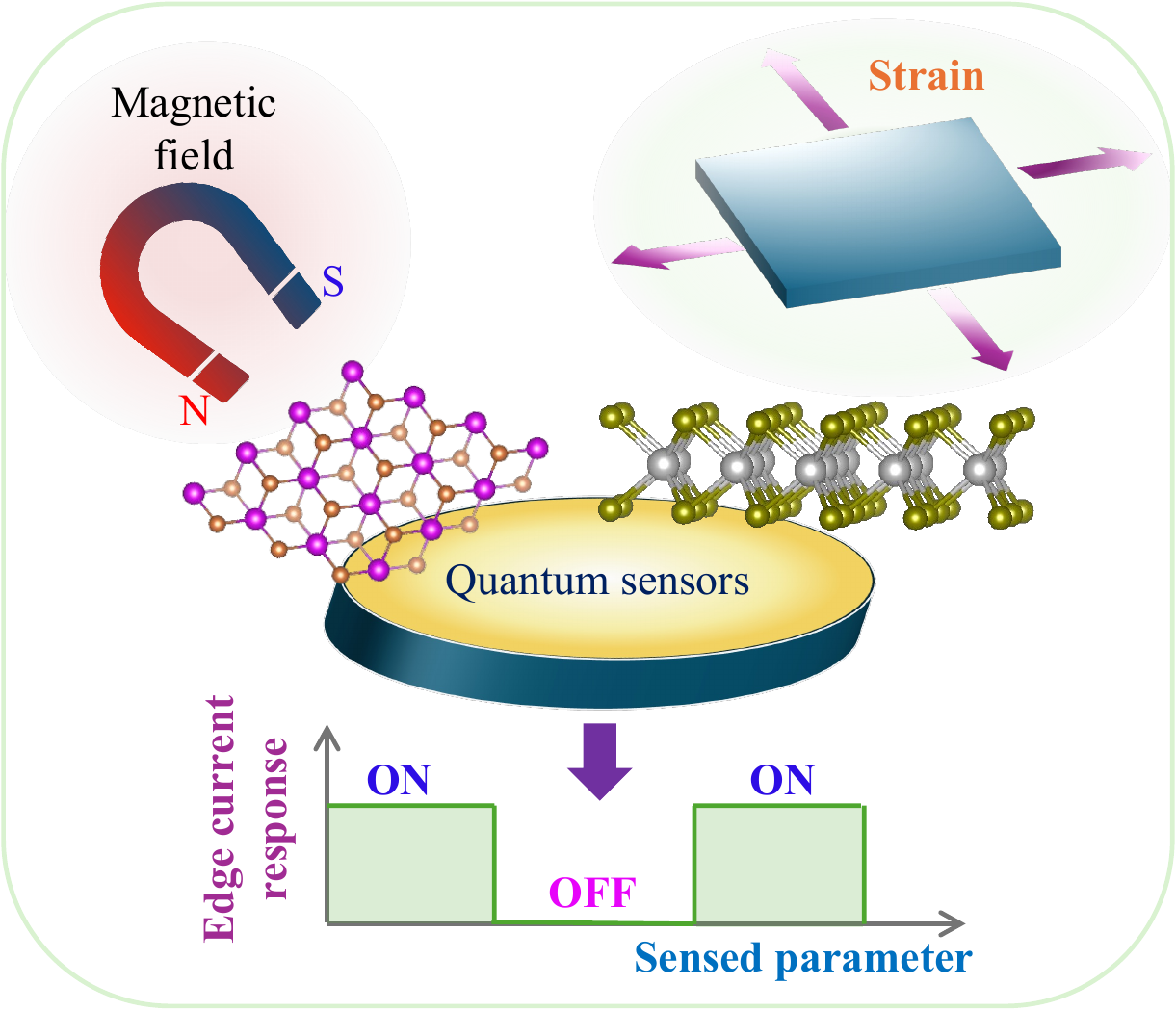}
\caption{Schematic illustration of potential applications of 2D vdW TIs in quantum sensing. Strain-driven topological phase transitions can be exploited for quantum strain sensors, while time-reversal-symmetry-protected edge states can provide a platform for designing highly sensitive magnetic field sensors. }
\label{fig:application}
\end{figure}

\subsection*{Promising future directions}

\begin{description}[leftmargin=0cm, labelwidth=5cm, labelsep=0.5em, align=left, style=nextline]

\item[High-throughput theoretical modeling]
Given the limited material landscape for 2D TIs with tunable topological phases, developing high-throughput, machine learning (ML)-based screening approaches could accelerate the identification of promising candidates without performing exhaustive first-principles calculations. One possible strategy is to construct simplified model Hamiltonians for representative systems, extract key parameters governing the topological phase transitions, and use these as input features to screen large materials databases. Such approaches can efficiently narrow down potential candidates for experimental and computational verification.

\item[Exploring novel material classes]

Exploring exotic materials with unique quantum properties may enable novel and higher-order topological phases in vdW heterostructures, even if the individual components are topologically trivial. Janus materials (e.g., TMDs, MXenes) and electrides provide promising platforms in this direction, and developing design principles for higher-order TIs in such heterostructures represents a promising direction for future research.

\item[Combining multiple design strategies]
While we have highlighted four primary approaches for designing tunable phases, combining these strategies could unlock further possibilities. For example, introducing light–matter interactions in moiré heterostructures may modulate interlayer coupling, stacking configurations, or twist angles, giving rise to spatially tunable non-equilibrium topological properties. Such hybrid approaches could create richer topological phenomena than achievable with a single design principle.

\item[Application-oriented opportunities]
Application-driven architectures based on 2D TIs offer promising routes toward novel device concepts~\cite{gilbert_Communication_Phy_2021, tian_Materials_2017}. For example, moiré heterostructures with spatially tunable topological domains can host designer edge states distributed throughout a planar device, unlike conventional TIs where edge states are confined to sample boundaries~\cite{tong_Nature-phy_2017}.

Additionally, 2D TIs offer promising platforms for quantum sensing applications  (Figure~\ref{fig:application}). Strain-tunable topological phase transitions can be exploited for mechanical strain sensing, while time-reversal-symmetry–protected edge states make them suitable for nanoscale magnetic field sensing. Systematic exploration of these functionalities could significantly expand the technological impact of 2D TIs beyond conventional electronic and spintronic applications.
\end{description}

Overall, two-dimensional topological quantum materials have emerged as a compelling platform for next-generation technologies. Realizing their full potential will require overcoming key material and scalability challenges while broadening the landscape of viable systems. Progress along these directions could enable a transformative class of quantum materials with technologically impactful applications.

\section{Acknowledgements}
 A.B.~and S.S.~acknowledge support from the U.S.~Department of Energy, Office of Science, Office of Fusion Energy Sciences, Quantum Information Science program under Award No.~DE-SC-0020340. Authors also acknowledge support from the Furth Research Fund at the University of Rochester. Authors thank the Pittsburgh Supercomputer Center (Bridges2) supported by the Advanced Cyberinfrastructure Coordination Ecosystem: Services \& Support (ACCESS) program, which is supported by National Science Foundation grants \#2138259, \#2138286, \#2138307, \#2137603, and \#2138296.  Authors thank Daniel Kaplan for fruitful discussions.

\bibliography{bibfile}
\end{document}